\documentclass[11pt]{article}
\usepackage{SIunits} 
\usepackage[margin=1in]{geometry} 
\usepackage{subfigure} 
\usepackage{graphicx} 
\usepackage{amsmath} 
\usepackage[format=hang,font=small,labelfont=bf]{caption} 
\usepackage[english]{babel} 
\usepackage{dirtytalk} 
\usepackage{nicefrac} 
\usepackage[hidelinks]{hyperref}
\usepackage[capitalise]{cleveref} 
\Crefname{equation}{Equation}{Equations}
\Crefname{figure}{Figure}{Figures}
\Crefname{tabular}{Table}{Tables}

\def\beq{\begin{equation}}
\def\eeq#1{\label{#1}\end{equation}}
\def\eeqn{\end{equation}}

\def\beqa{\begin{eqnarray}}
\def\eeqa#1{\label{#1}\end{eqnarray}}
\def\eeqan{\end{eqnarray}}

\let\bar=\overbar

\def\Dslash{\not{\hbox{\kern-4pt $D$}}}
\def\dslash{\not{\hbox{\kern-2pt $\del$}}}

\def\msb{{\bar{\ssstyle M \kern -1pt S}}}

\def\Title#1{\begin{center} {\Large {\bf #1} } \end{center}}
\def\Author#1{\begin{center} {\normalsize {\sc #1} } \end{center}}
\def\Institution#1{\begin{center} {\normalsize {\it #1} } \end{center}}
\def\Collaboration#1{\begin{center} {\normalsize #1 } \end{center}}
\def\Abstract#1{\noindent {\normalsize {\bf Abstract:} {\normalfont #1}}}
\def\Conference{\vspace{4mm}\begin{raggedright} {\normalsize {\it Talk presented at the 2019 Meeting of the Division of Particles and Fields of the American Physical Society (DPF2019), July 29--August 2, 2019, Northeastern University, Boston, C1907293.} } \end{raggedright}\vspace{4mm}}

\begin{document}

%
%

\Title{The Fermilab Muon $g-2$ straw tracking detectors, internal tracker alignment, and the muon EDM measurement}

\Author{Gleb Lukicov\footnote{e-mail: g.lukicov@ucl.ac.uk}}

\Collaboration{on behalf of the Fermilab Muon $g-2$ Collaboration\footnote{https://muon-g-2.fnal.gov/collaboration.html}}

\Institution{Department of Physics and Astronomy, University College London,\\ London, WC1E 6BT, United Kingdom}

\Abstract{The $g-2$ experiment successfully completed its second data taking period (Run-2) between March and July 2019, with the analysis of Run-1 and Run-2 data currently proceeding. The final goal of the experiment is to determine the muon magnetic anomaly, $a_{\mu}=\frac{g-2}{2}$, to a precision of 140 ppb. Essential to reducing the systematic uncertainty on $a_{\mu}$ through measurements of the muon beam profile are the in-vacuum straw tracking detectors. A crucial prerequisite to obtaining accurate distributions of the beam profile is the internal alignment of the trackers, which is described here, along with the resulting beam distributions. This paper also discusses an additional measurement that will be made using the trackers: setting a new limit on the electric dipole moment (EDM) of the muon, producing a world-leading measurement.}

\Conference 

\section{Introduction}
High intensity muon experiments allow for stringent tests of the Standard Model (SM). The current leading measurement \cite{BNL_AMM} of the muon magnetic anomaly, $a_{\mu}$, at Brookhaven National Laboratory (BNL) yielded a discrepancy between the theoretically predicted and experimentally measured values of more than $3\sigma$: a possible indication of New Physics (NP). The Fermilab $g-2$ experiment \cite{FNAL_TDR} will determine $a_{\mu}$ to a precision of 140 ppb, sufficient to establish the presence of NP to $7\sigma$ should the same central value be measured.

The muon has an intrinsic magnetic dipole moment, $\boldsymbol{\mu}$, that is coupled to its spin \cite{LDM}, $\boldsymbol{s}$, by the $g$-factor $g_{\mu}$
\begin{equation}
    \boldsymbol{\mu}=g_\mu\left(\frac{e}{2m_{\mu}}\right)\boldsymbol{s}.
    \label{eq:mdm}
\end{equation}
The Dirac equation predicts the value of $g_{\mu}$ to be exactly equal to 2, however, additional radiative corrections cause it to be slightly larger than 2. This difference, $a_{\mu}$, is defined as
\begin{equation}
a_{\mu}=\frac{g_{\mu}-2}{2}.
\label{eq:AMM}
\end{equation}

\begin{figure}[htb]
\centering
\subfigure[]{\includegraphics[height=1.3in]{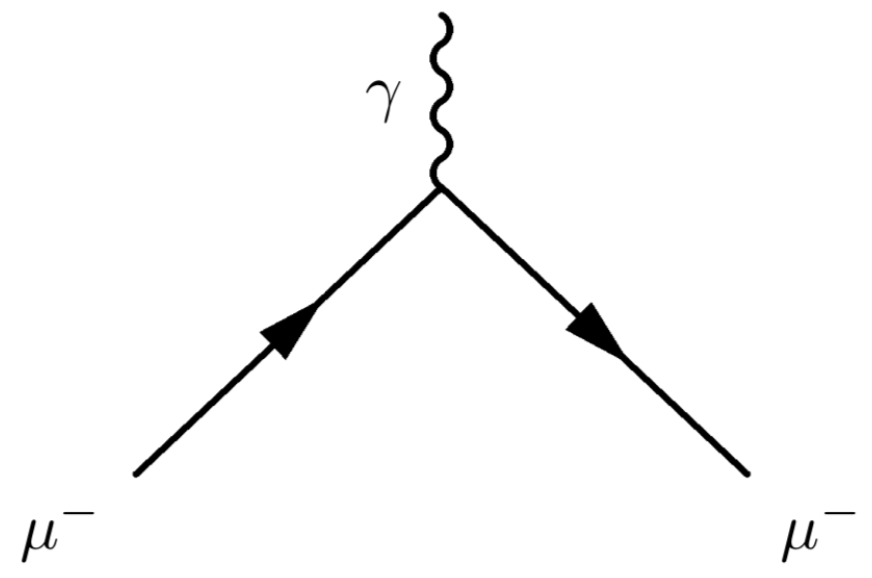}}
\subfigure[]{\includegraphics[height=1.3in]{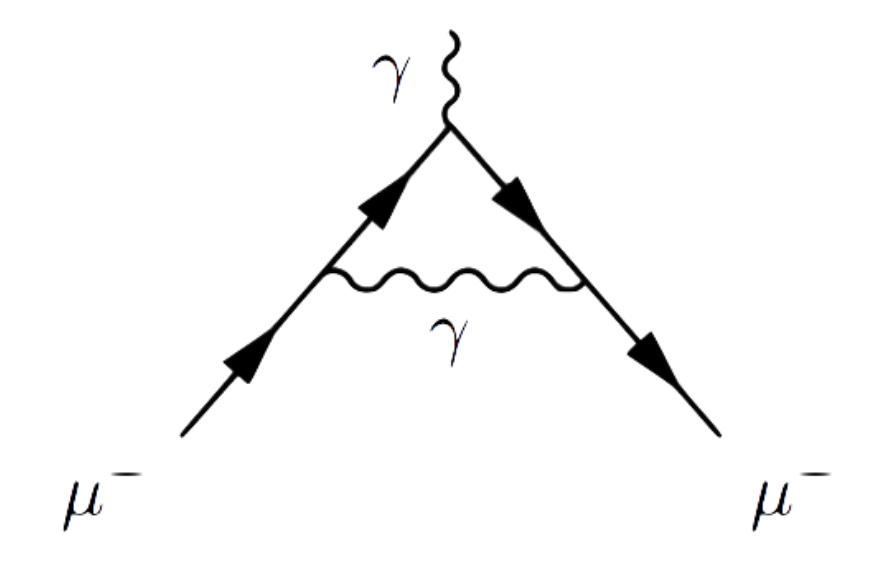}}
\subfigure[]{\includegraphics[height=1.3in]{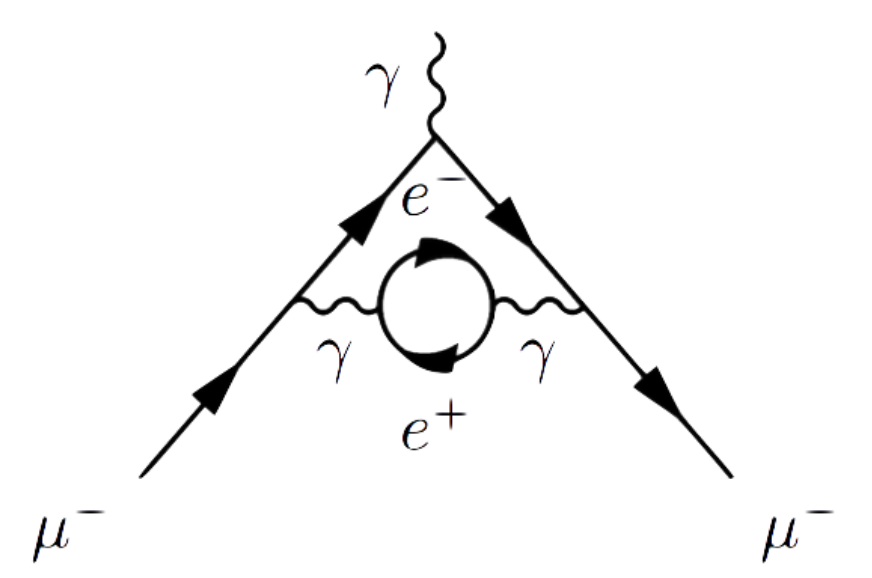}} \\
\subfigure[]{\includegraphics[height=1.3in]{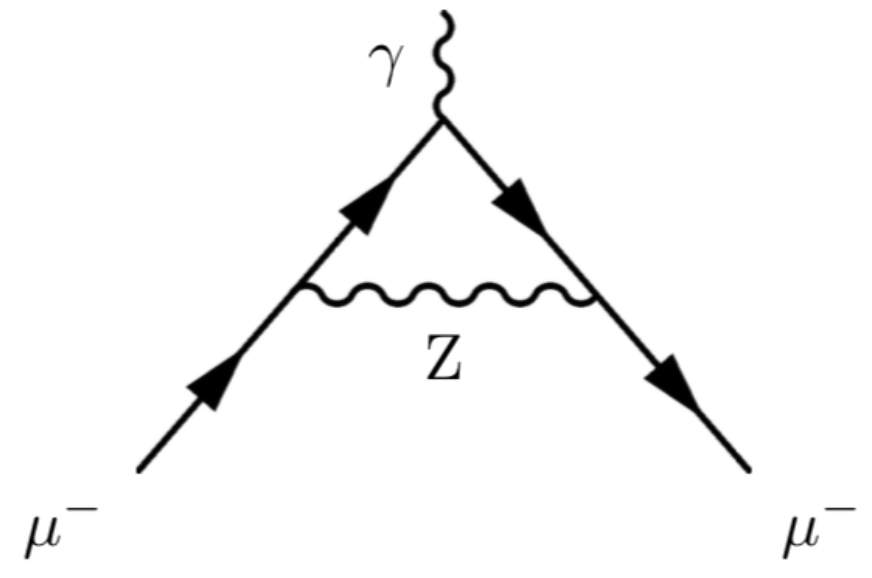}}
\subfigure[]{\includegraphics[height=1.3in]{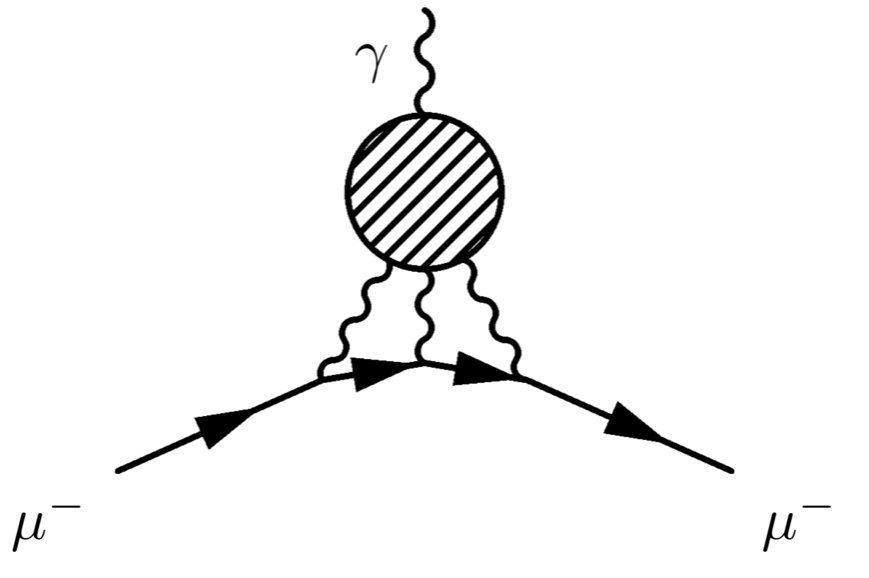}}
\subfigure[]{\includegraphics[height=1.3in]{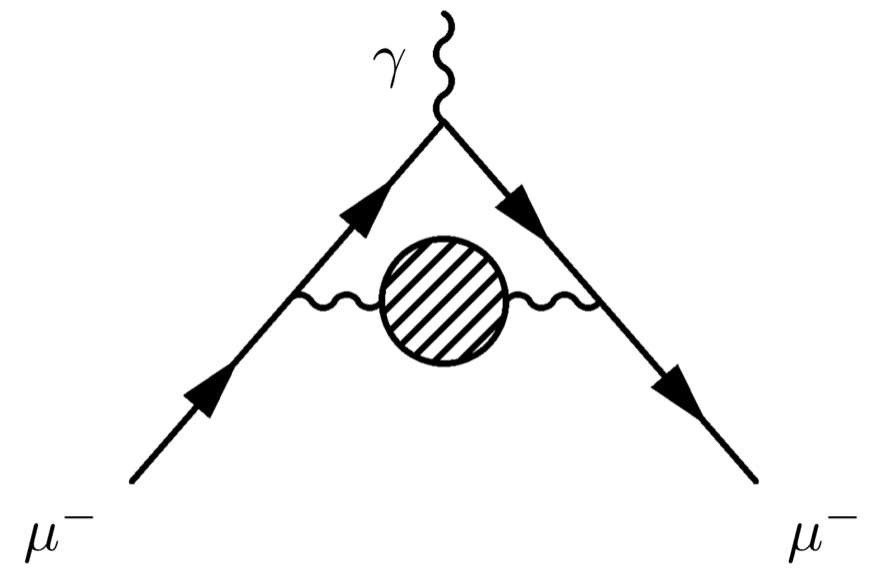}}
\caption{Examples of SM contributions to $a_{\mu}$: a) Dirac b) Schwinger c) higher order QED d) electroweak e) hadronic light-by-light f) hadronic vacuum polarisation.}
\label{fig:feynam}
\end{figure}

\begin{figure}[!htb]
\centering
\includegraphics[height=1.7in]{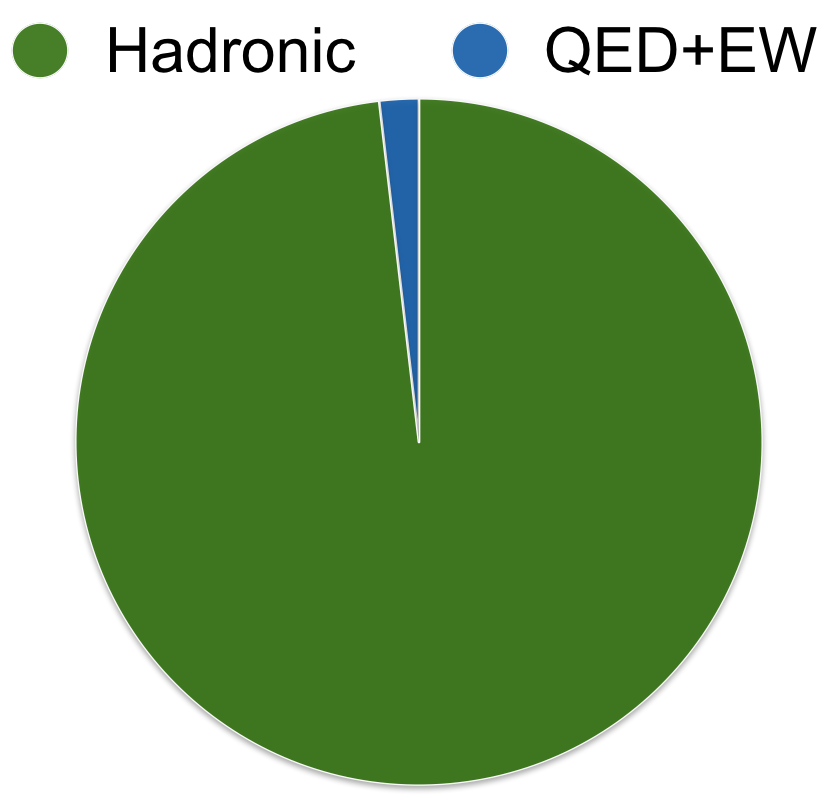}
\caption{The uncertainty on the theoretical prediction of $a_{\mu}$ is dominated by the hadronic interactions.}
\label{fig:error}
\end{figure}

The theoretical calculation of $a_{\mu}$ \cite{KNT} is of a similar level of precision as the experimentally measured value, and combines contributions from electromagnetic (QED), electroweak (EW), and hadronic interactions, as shown in \cref{fig:feynam}. The uncertainty on the calculation is dominated by the hadronic contributions as shown in \cref{fig:error}. A new theoretical result with reduced uncertainties is expected on the timescale of the final experimental result \cite{Theory}.

\section{Methodology of the \texorpdfstring{$g-2$}\xspace~experiment at Fermilab}
A reduction in the experimental uncertainty on $a_{\mu}$ will be achieved via reductions in the statistical uncertainty due to the use of the Fermilab muon beam \cite{Diktys}, and the systematic uncertainty due to improvements in technology such as: in-vacuum tracking detectors \cite{Tom, James}, segmented calorimeters~\cite{Jarek}, better field uniformity \cite{Mark}, and a laser calibration system \cite{Anastasi}, as well as analysis techniques.

Achieving the experimental goal requires a dataset containing $1.5 \times 10^{11}$ positrons with energy above 1.8 GeV. The momentum of the injected muons is 3.09 GeV: the so-called \say{magic momentum} at which the effect of the motional magnetic field \cite{FNAL_TDR} is zero. To perform the measurement of $a_{\mu}$, two measurable quantities are of interest: the anomalous precession frequency, $\boldsymbol{\omega_a}$, and the applied magnetic field, $\boldsymbol{B}$. $\boldsymbol{\omega_a}$ is defined as the difference between the spin and cyclotron frequencies. In the case of a uniform magnetic field and negligible betatron oscillations, the relationship between $a_{\mu}$, $\boldsymbol{\omega_a}$, and $\boldsymbol{B}$ can be written as follows
\vspace{-0.1cm}
\begin{equation}
\boldsymbol{\omega_a}=a_{\mu}\frac{e}{m_{\mu}}\boldsymbol{B}=\left(\frac{g-2}{2}\right)\frac{e}{m_{\mu}}\boldsymbol{B}.
\label{eq:mub}
\end{equation}
\clearpage
\begin{figure}[htb]
\centering
\subfigure[]{\includegraphics[width=0.47\linewidth]{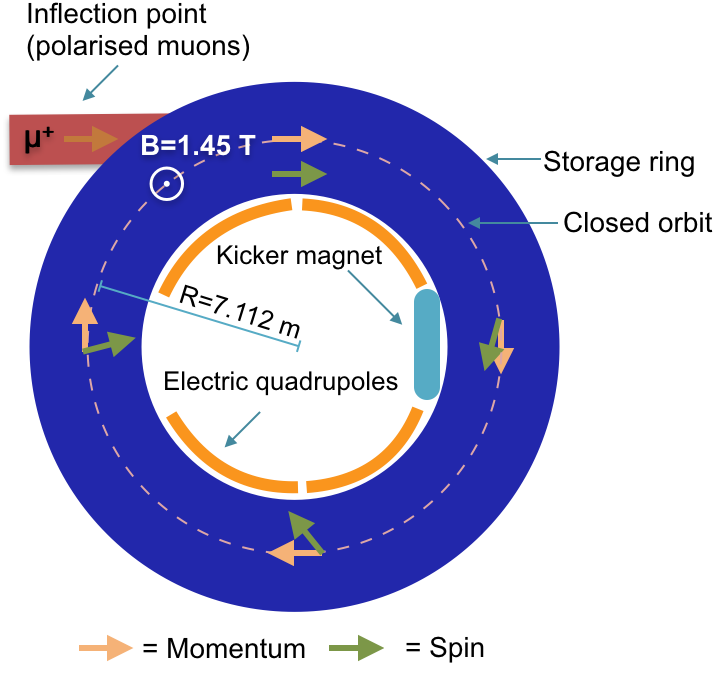}\label{fig:ring}}
\subfigure[]{\includegraphics[width=0.47\linewidth]{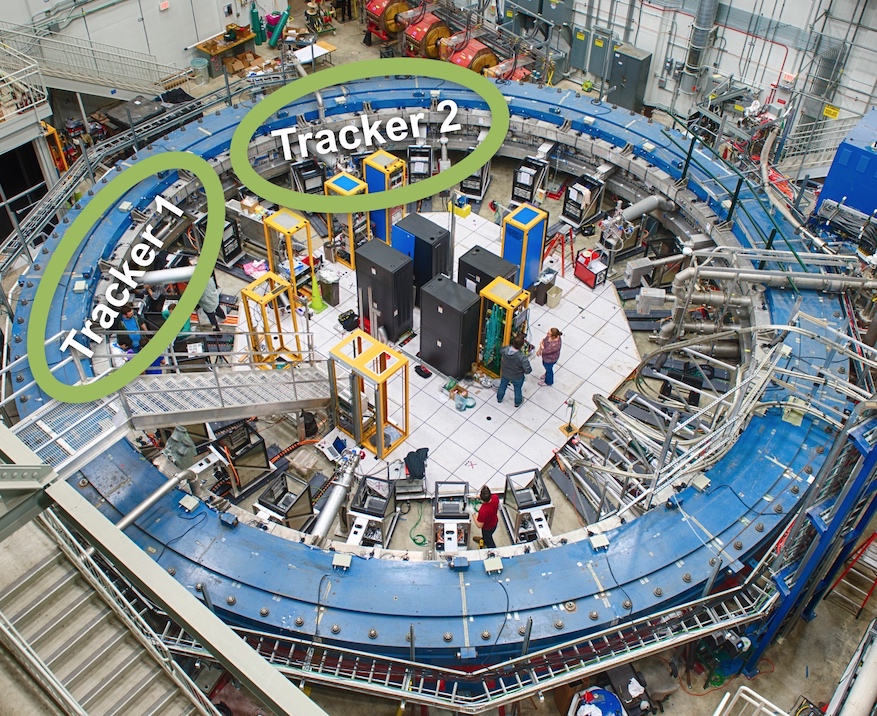}\label{fig:ring_pic}}
\vspace{-0.35cm}
\caption{The $g-2$ ring : a) Superconducting magnets provide a uniform radial magnetic dipole field of 1.45 T. The muon beam enters the storage ring via the inflector. The direction of the beam is clockwise in this orientation. b) The image of the ring with the two tracker stations indicated. The 24 calorimeter readout electronic frames (shown in black) are seen distributed uniformly on the inside of the ring.}
\end{figure}
\vspace{-0.25cm}
Longitudinally polarised muons are injected into the storage ring, as shown in \cref{fig:ring}, where they follow a circular orbit in the magnetic dipole field and are vertically focused using electrostatic quadrupoles. The muon beam enters the ring though the inflector magnet, designed to cancel out the dipole magnetic field, at a slightly larger radius than the ideal orbit. The kicker magnets are used to deflect the beam onto the correct orbit.

\begin{figure}[htb]
\centering
\subfigure[]{\raisebox{10mm}{\includegraphics[width=0.49\linewidth]{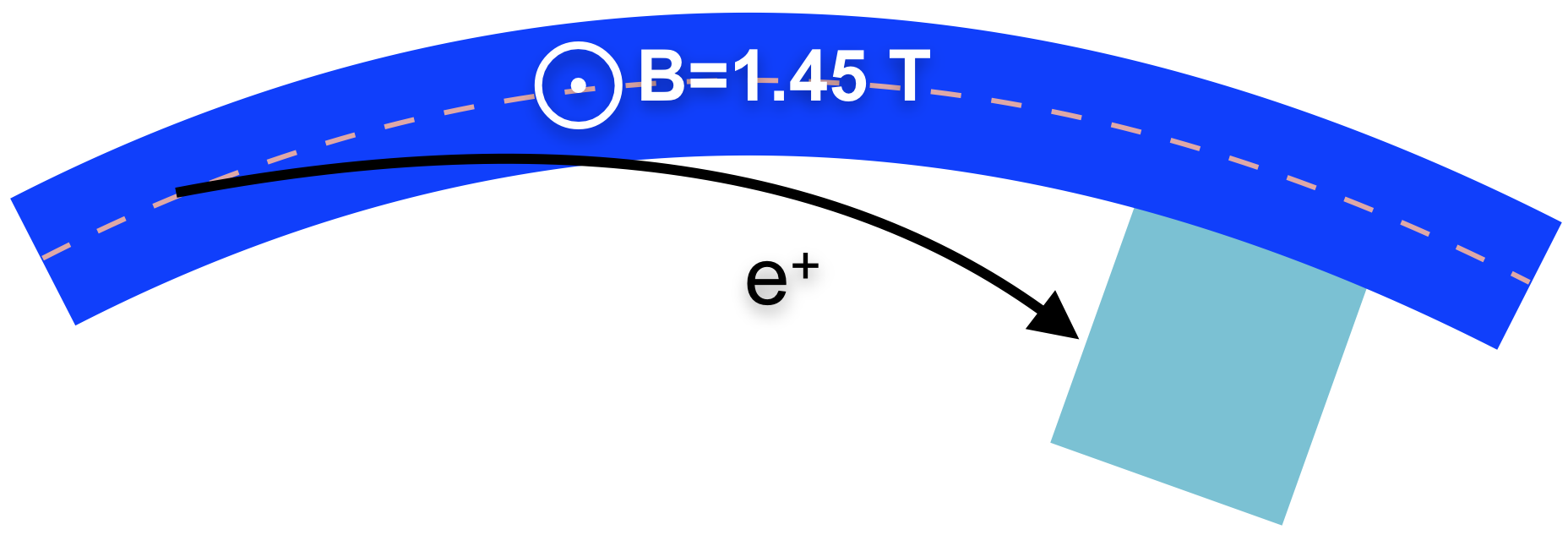}\label{fig:calo}}}
\subfigure[]{\includegraphics[width=0.49\linewidth]{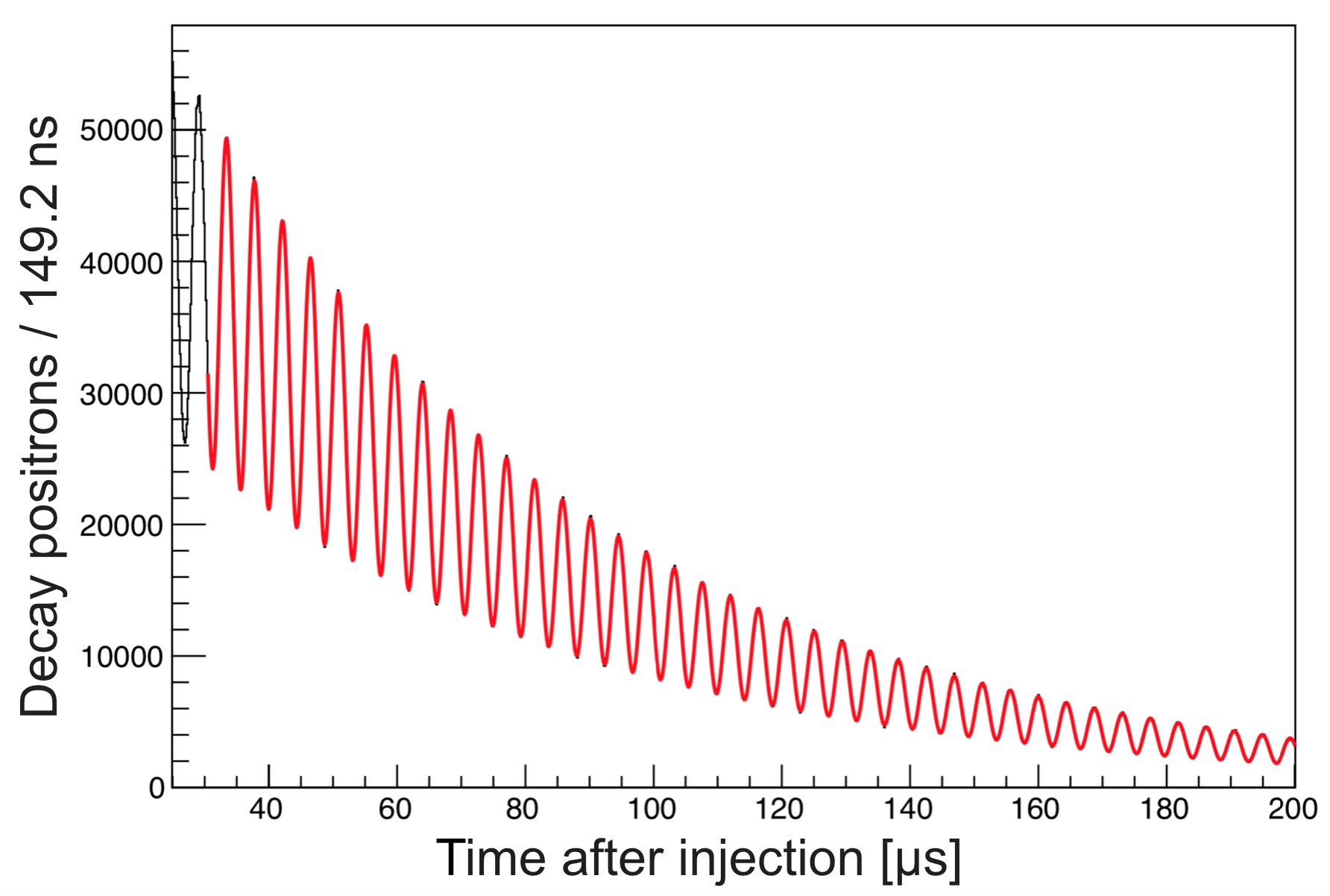}\label{fig:wiggle}}
\caption{Measurement of $\boldsymbol{\omega_a}$: a) Decay positron entering one of the 24 calorimeters. b) $\boldsymbol{\omega_a}$ is extracted from a fit to calorimeter events after $30~\micro\second$, when the beam is stable.}
\end{figure}

Since $a_{\mu} > 0 $, the muon spin precesses faster than the momentum vector in the storage ring. Due to the $V-A$ structure of the weak interaction, the highest energy decay positrons are preferentially emitted along the direction of the muon spin. The decay positrons curl inwards, where they can interact with one of the 24 calorimeters placed around the interior of the storage ring. This is shown in \cref{fig:calo}. High energy positron events are then histogrammed as a function of time to extract $\boldsymbol{\omega_a}$, as shown in \cref{fig:wiggle}.

The magnetic field is determined through a frequency measurement: Larmor frequency of a free proton, $\omega_p$. It is measured using Nuclear Magnetic Resonance (NMR) probes to establish the magnetic flux density, $|\boldsymbol{B}|$
\begin{equation}
\omega_p = \gamma_p|\boldsymbol{B}|,
\end{equation}
where $\gamma_p$ is the free proton gyromagnetic moment ratio \cite{CODATA}. $a_{\mu}$ can then be expressed as a function of the two experimentally measured frequencies and well-determined ratios
\begin{equation}
a_{\mu} = \frac{\nicefrac{\omega_a}{\omega_p}}{\nicefrac{\mu_{\mu}}{\mu_p}-\nicefrac{\omega_a}{\omega_p}}=\frac{\omega_a}{\omega_p}\left(\frac{g_e}{2}\right)\left(\frac{m_{\mu}}{m_e}\right)\left(\frac{\mu_p}{\mu_e}\right),
\end{equation}
where $g_e$, $\nicefrac{\mu_p}{\mu_e}$, and $\nicefrac{m_{\mu}}{m_e}$ are known \cite{CODATA} with 0.00026 ppb, 3.0 ppb, 22 ppb in uncertainty, respectively.
\clearpage

Beam dynamics effects, such as the momentum spread ($\Delta \boldsymbol{p}$) and Coherent Betatron Oscillations (CBO) \cite{CBO}, contribute significantly to the $\boldsymbol{\omega_a}$ systematic uncertainty budget. Since the calorimeter acceptance \cite{Aaron} of the decay positrons depends on the radius of the muon at the point of decay, the CBO of the stored beam can produce an amplitude modulation in the observed positron time spectrum. The tracking detectors are used to measure and monitor the CBO. Moreover, the largest single systematic uncertainty (40 ppb) associated with the calorimeters is pileup, which occurs when, for example, two low energy positrons deposit energy in the same crystal close together in time ($\sim 5$ ns). The tracking detectors can be used to investigate pileup in the adjacent calorimeters, as they can reconstruct independent trajectories of the positrons. 

\section{Tracker overview}
\begin{figure}[!ht]
    \centering
    \subfigure[]{\raisebox{10mm}{\includegraphics[width=.46\linewidth]{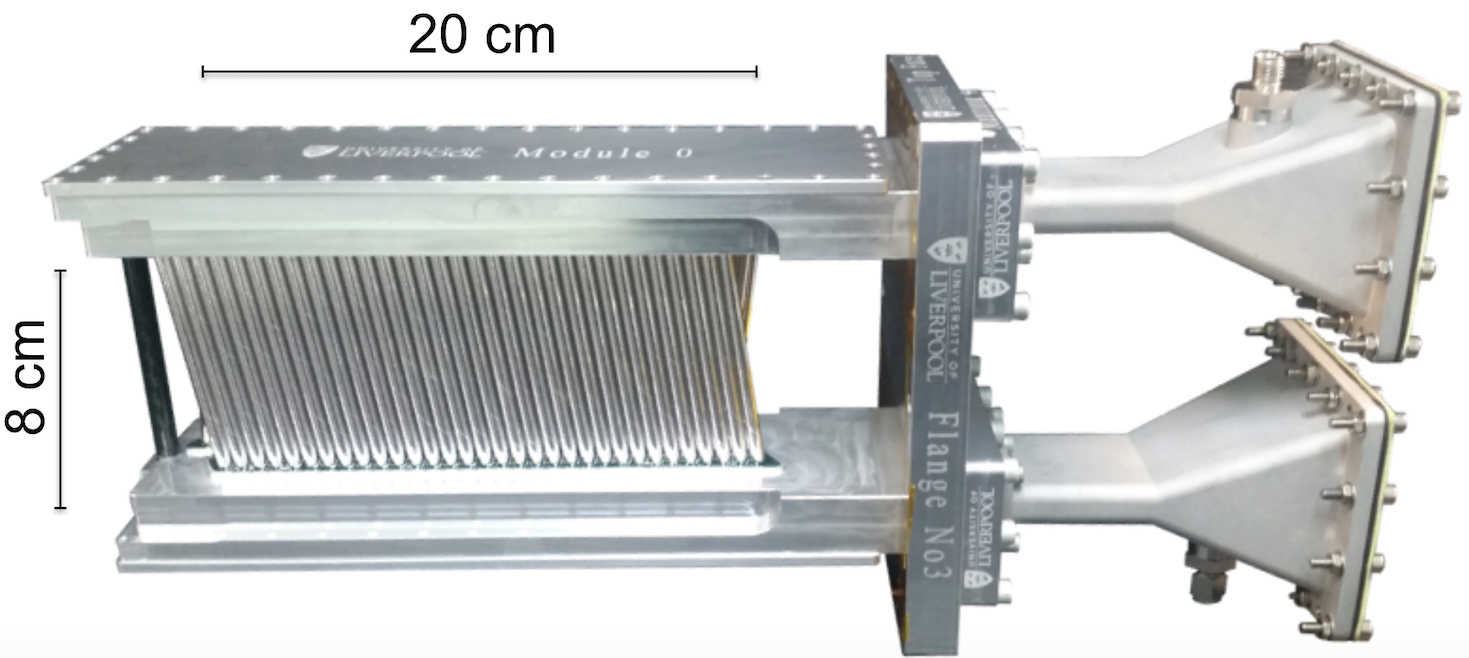} \label{fig:tracker_photo}}}
    \subfigure[]{\includegraphics[width=.46\linewidth]{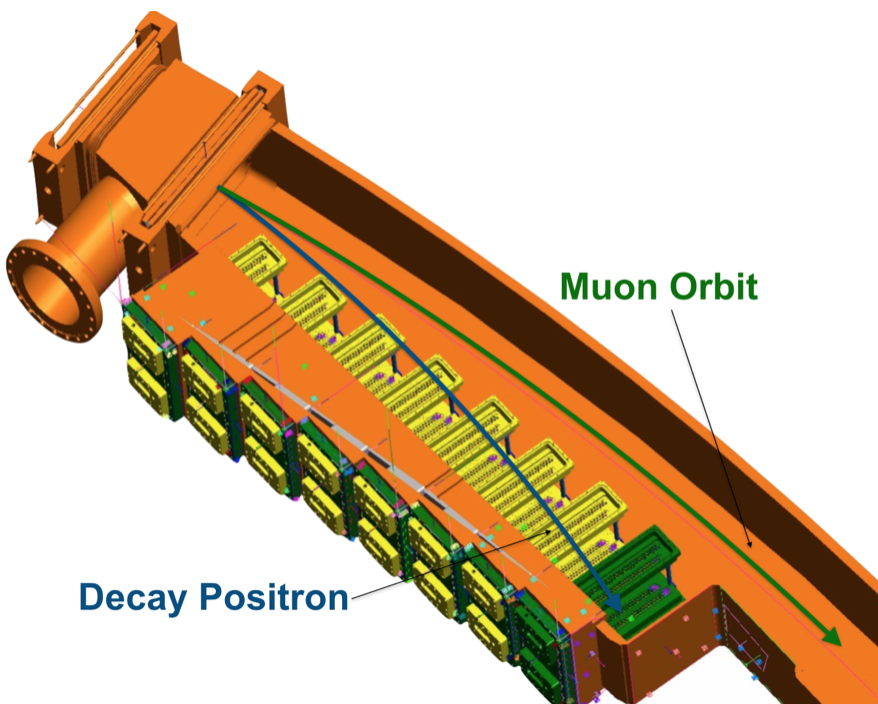} \label{fig:station}}
     \caption{Tracking detector: a) A single tracker module with one of the four rows of 32 straws, which are held between two manifolds. A straw (length=100 mm, diameter=5 mm) is an ionisation chamber filled with 50:50 Ar:Ethane, with a central anode wire at +1.6 kV. The active tracking region is inside the storage ring vacuum ($10^{-9}$ atm). b) Rendering of a decay positron trajectory passing through the tracker station before hitting the calorimeter.}
\end{figure}

The tracking detector \cite{Tom}, shown in \cref{fig:tracker_photo}, measures the trajectory of the positrons from the $\mu^+$ decay in the storage ring. Each tracker module consists of four layers of 32 straws oriented at $\pm7.5^{\circ}$ to the vertical. Each aluminised mylar straw is $15~\micro\metre$ thick, and is held at 1~atm pressure. Each straw is filled with a 50:50 Ar:Ethane mixture and contains a central wire that is held at a $+1.6$~kV potential. The modules are inside the vacuum of $10^{-10}$ atm. Eight tracker modules make up a tracker station, shown in \cref{fig:station}, with the two stations located in front of two calorimeters, as indicated in \cref{fig:ring_pic}. 
The primary aim of the tracker is to reduce the systematic uncertainty on $\boldsymbol{\omega_a}$ via measurements and assessments of the muon beam profile, beam dynamics, positron pile-up, and calorimeter gains.

\section{Tracking and extrapolation}
\begin{figure}[!ht]
    \centering
    \subfigure[]{\includegraphics[width=.46\linewidth]{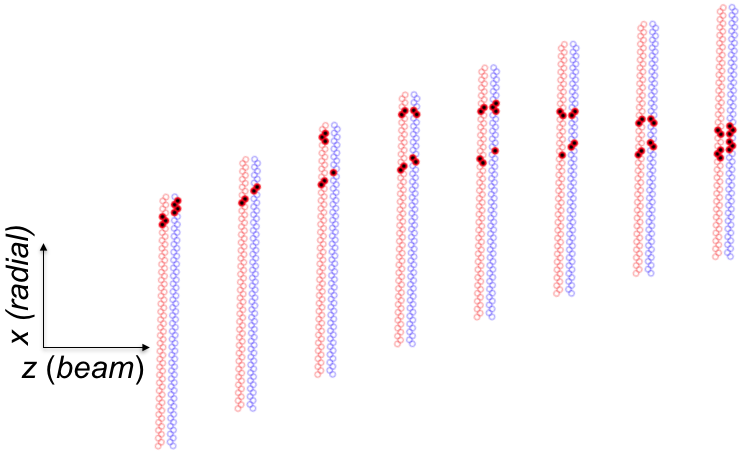} \label{fig:hits}}
    \subfigure[]{\includegraphics[width=.46\linewidth]{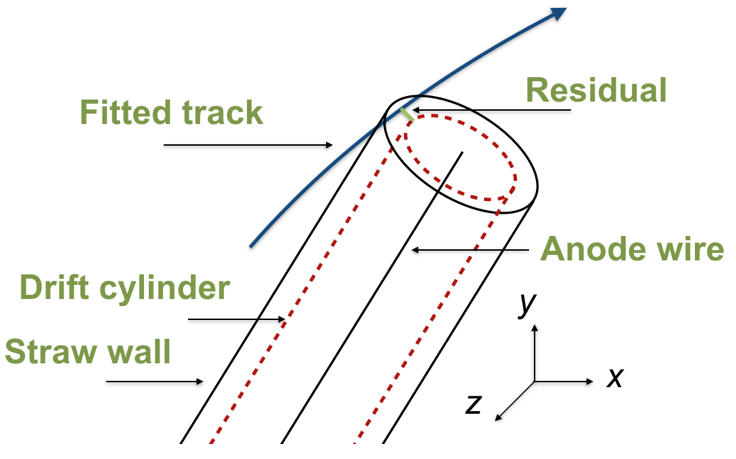} \label{fig:straw}}
    \caption{The track fitting technique: a) Hits from two tracks close in time  b) A fitted track passing though a single straw. A residual between the fitted track and the measurement is indicated. The measurement in the straw is a hit time, which is converted into a drift radius to produce a drift cylinder around the central anode wire.}
\end{figure}

\begin{figure}[!ht]
    \centering
    \includegraphics[width=.76\linewidth]{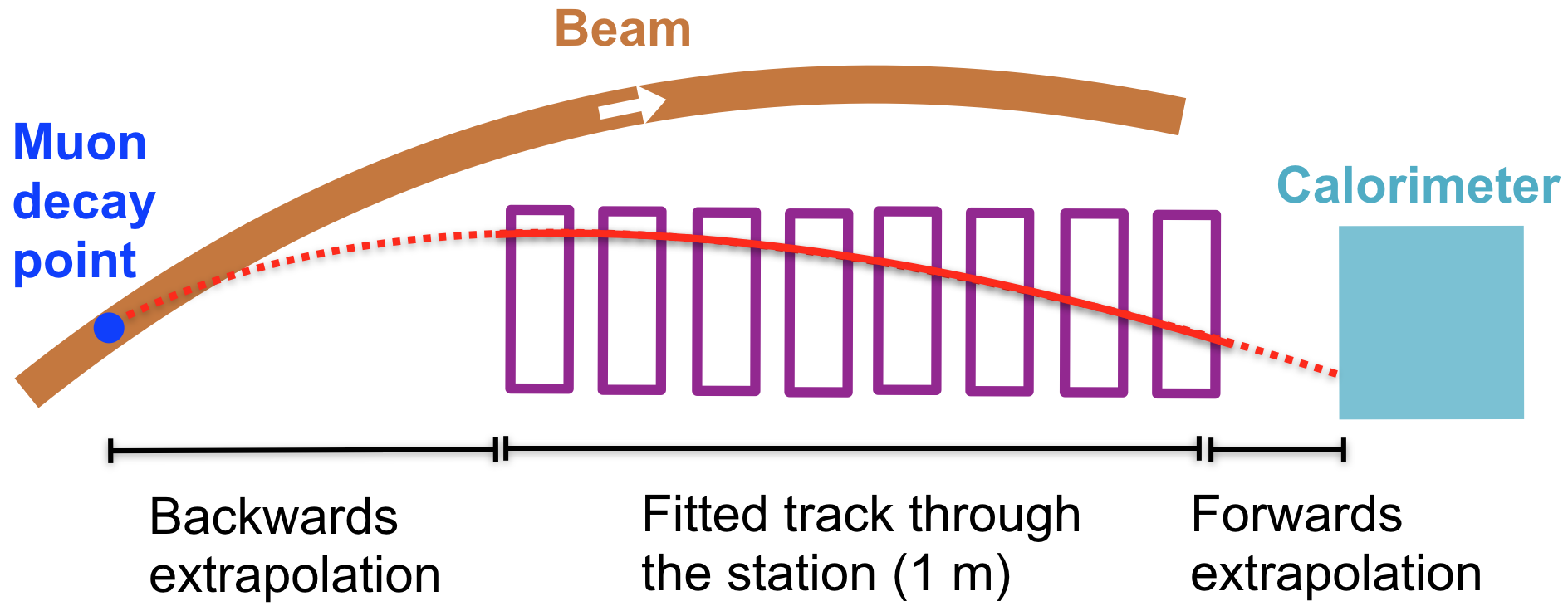} 
    \caption{The track extrapolation technique: a fitted track is extrapolated backwards to the muon decay point, and forwards into the calorimeter.}
    \label{fig:extrap}
\end{figure}

Track reconstruction is implemented with the \verb!GEANE! framework \cite{Nick}, which incorporates geometry, material, and field, utilising transport and error matrices for particle propagation though the straw planes, as shown in \cref{fig:hits}.   
The resulting fitted tracks are then extrapolated \cite{Saskia} back to the most probable muon decay point, as shown in \cref{fig:extrap}, using a \textit{Runge-Kutta} algorithm that propagates the tracks through the varying magnetic field, until point of radial tangency is reached. Tracks that pass through any material during the extrapolation are rejected. 

\begin{figure}[!ht]
    \centering
    \subfigure[]{\raisebox{3.5mm}{\includegraphics[width=.52\linewidth]{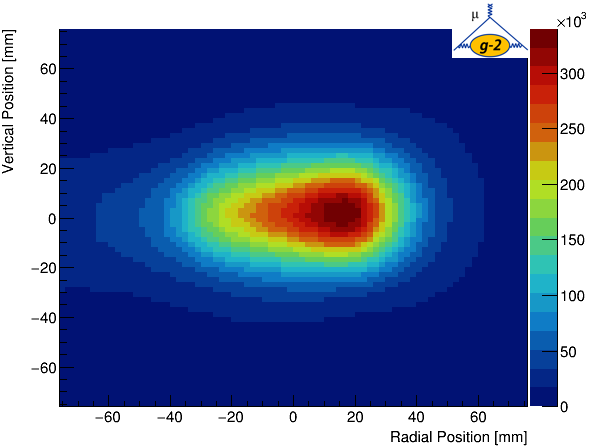} \label{fig:beam}}}
    \subfigure[]{\includegraphics[width=.45\linewidth]{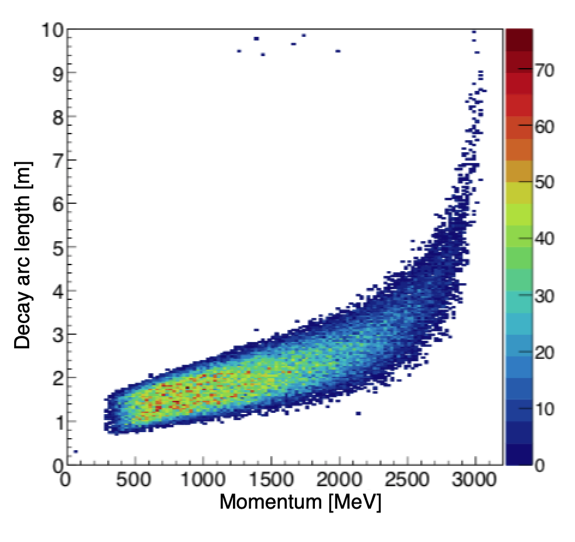} \label{fig:arc}}
    \caption{Backwards extrapolation: a) Reconstructed beam profile from tracks that have been extrapolated back to their decay position. b) Reconstructed decay arc length as a function of track momentum.}
\end{figure}
\begin{figure}[!ht]
    \centering
    \subfigure[]{\includegraphics[width=.48\linewidth]{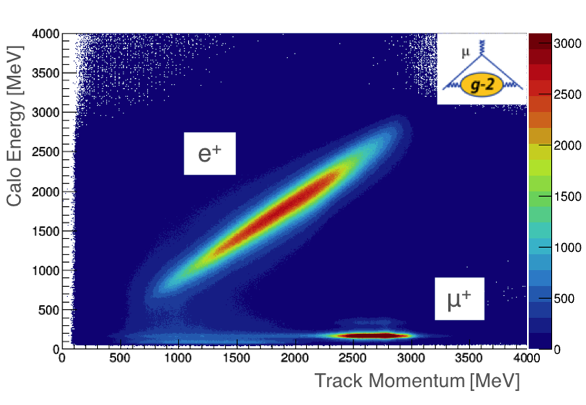} \label{fig:ep}}
    \subfigure[]{\includegraphics[width=.49\linewidth]{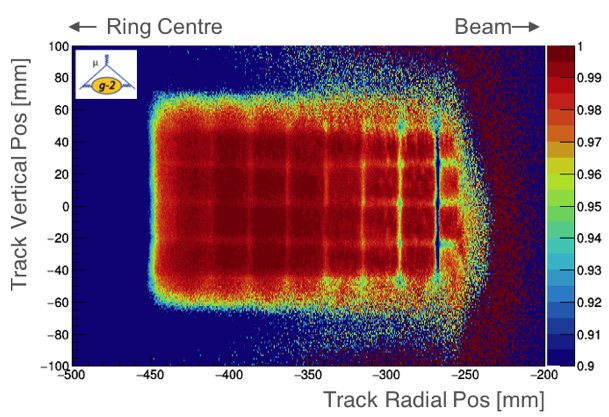} \label{fig:face}}
    \vspace{-0.1cm}
    \caption{Forwards extrapolation: Tracks and calorimeter clusters are matched based on their time proximity: a) Comparing the measured momentum from the trackers with the energy from the calorimeters shows two distinct populations - positrons that are mostly contained in the calorimeter, and high momentum lost muons which only deposit a small amount of energy in the calorimeter.   b) Extrapolated tracks to the front face of the calorimeter. The number of matched tracks is divided by the total number of tracks to show the efficiency as a function of position. Nearly all the missing calorimeter hits resemble lost muons. The efficiency decreases in the gaps between the crystals where a lost muon might split its small energy deposition between the two crystals, neither of which goes above the threshold. This contains early times in the fill (\textless~$30~\micro\second$), so there are more lost muons than during the data-taking period, which artificially lowers the efficiency, but brings out the crystal structure.}
\end{figure}
\begin{figure}[!ht]
    \centering
    \includegraphics[width=.53\linewidth]{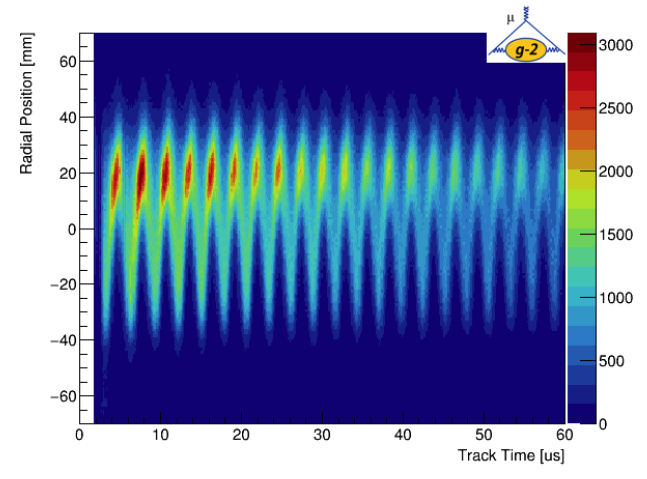} 
    \vspace{-0.1cm}
    \caption{Reconstructed radial position of the muon beam plotted against time. The oscillations are the radial betatron oscillations, known as the CBO.}
    \label{fig:cbo}
\end{figure}

The extrapolated tracks are used to measure the muon beam profile, as shown in \cref{fig:beam}, as well as to calculate the decay arc length, as shown in \cref{fig:arc}. The tracks can also be extrapolated forward to the calorimeter, enabling particle identification, as shown in \cref{fig:ep}, as well as to investigate the efficiency of matched calorimeter clusters and tracks, as shown in \cref{fig:face}.
Additionally, the radial position of the muon beam can be monitored as a function of time, as shown in \cref{fig:cbo}, in order to estimate the amplitude and frequency of the CBO. 

\section{Pitch correction and field convolution}
\begin{figure}[!ht]
    \centering
    \subfigure[]{\raisebox{6mm}{\includegraphics[width=.48\linewidth]{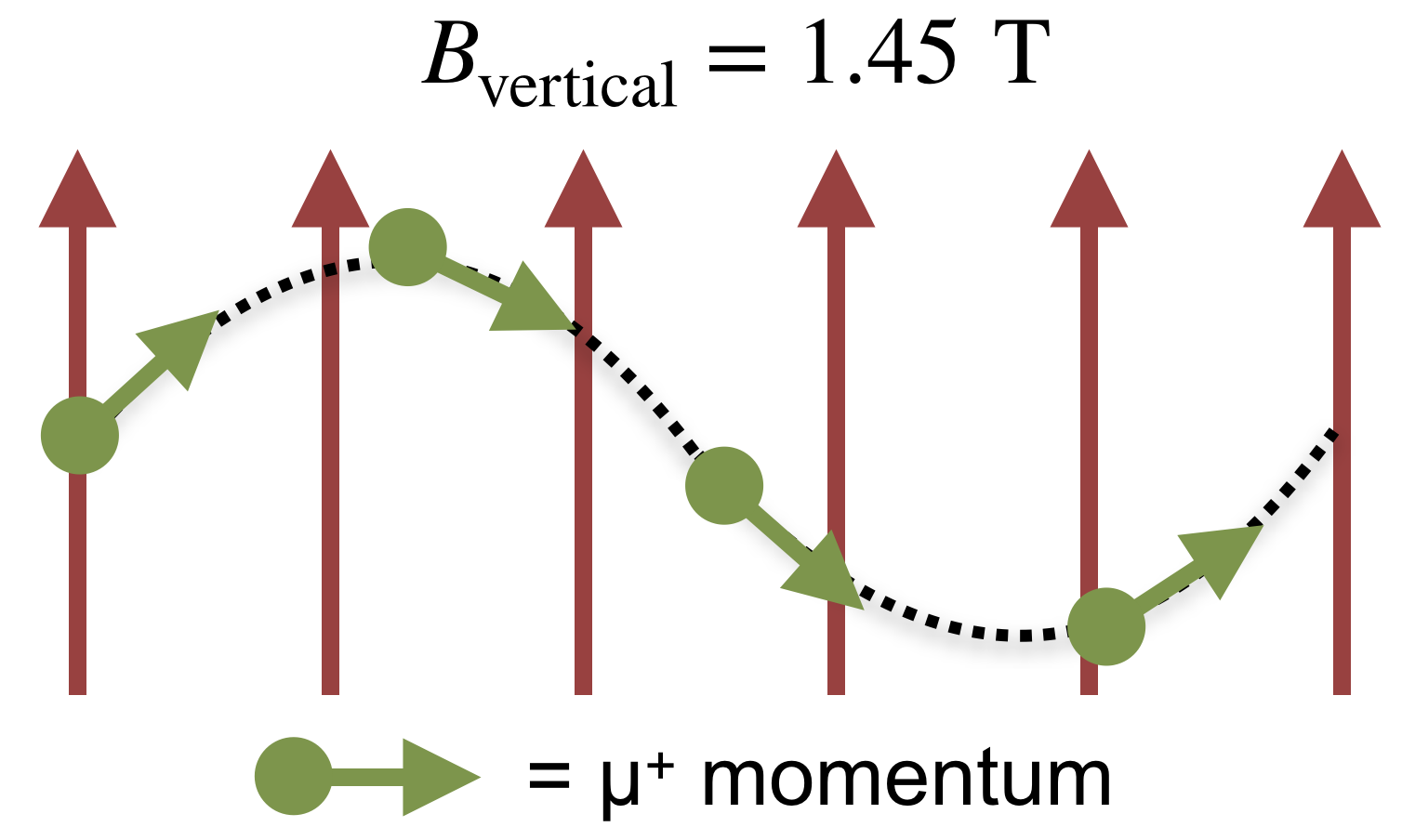} \label{fig:pitch}}}
    \subfigure[]{\includegraphics[width=.48\linewidth]{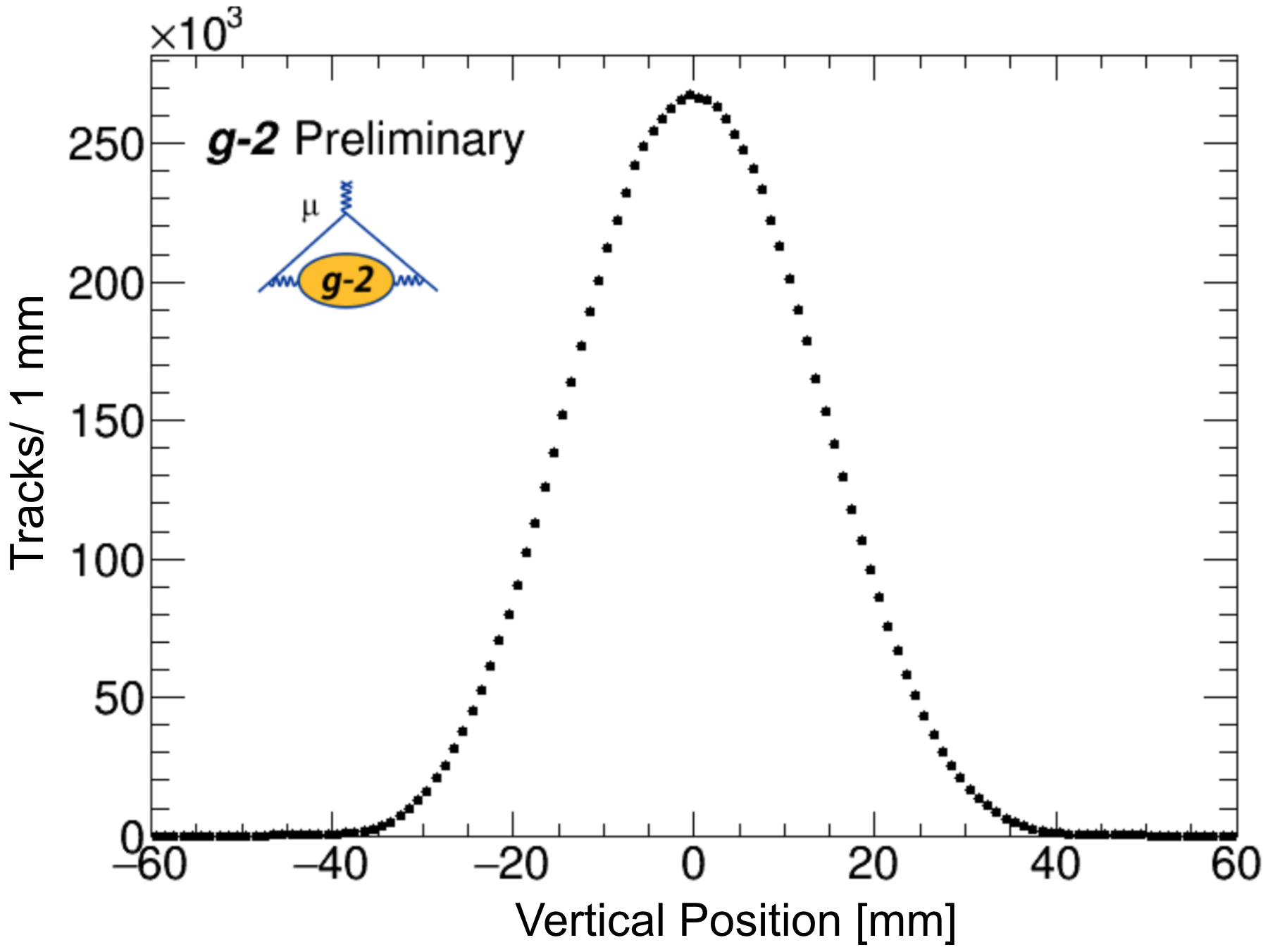} \label{fig:ver}}
    \vspace{-0.25cm}
    \caption{Vertical pitch: a) The up-and-down motion experienced by the muons due to the quadrupole electric field. b) Vertical width of the beam from the trackers.}
\end{figure}

The muons in the storage ring undergo a vertical motion due to the focusing from the electrostatic quadruples, as shown in \cref{fig:pitch}. This introduces an additional term to the expression of $\boldsymbol{\omega_a}$ in \cref{eq:mub} that needs to be incorporated into the analysis
\begin{equation}
    \boldsymbol{\omega_a}= \frac{e}{m_{\mu}} \bigg[ a_{\mu}\boldsymbol{B} - \frac{\gamma a_{\mu}}{\gamma +1}(\boldsymbol{B}\cdot \boldsymbol{\beta})\boldsymbol{\beta}  \bigg].
\end{equation}
This changes the measured spin precession frequency, and so needs to be corrected for, the so-called \textit{pitch correction}, which is proportional to the vertical width of the beam, $\sigma_{\mathrm{vertical}}$
\begin{equation}
    \frac{\Delta\omega_a}{\omega_a}  \propto \sigma_{\mathrm{vertical}}^{2}.
\end{equation}
The vertical width is measured by the trackers, as shown in \cref{fig:ver}, with the uncertainty on the correction within the uncertainty budget for this systematic effect \cite{FNAL_TDR}.

\begin{figure}[!ht]
    \centering
    \hspace*{1cm}\includegraphics[width=.5\linewidth]{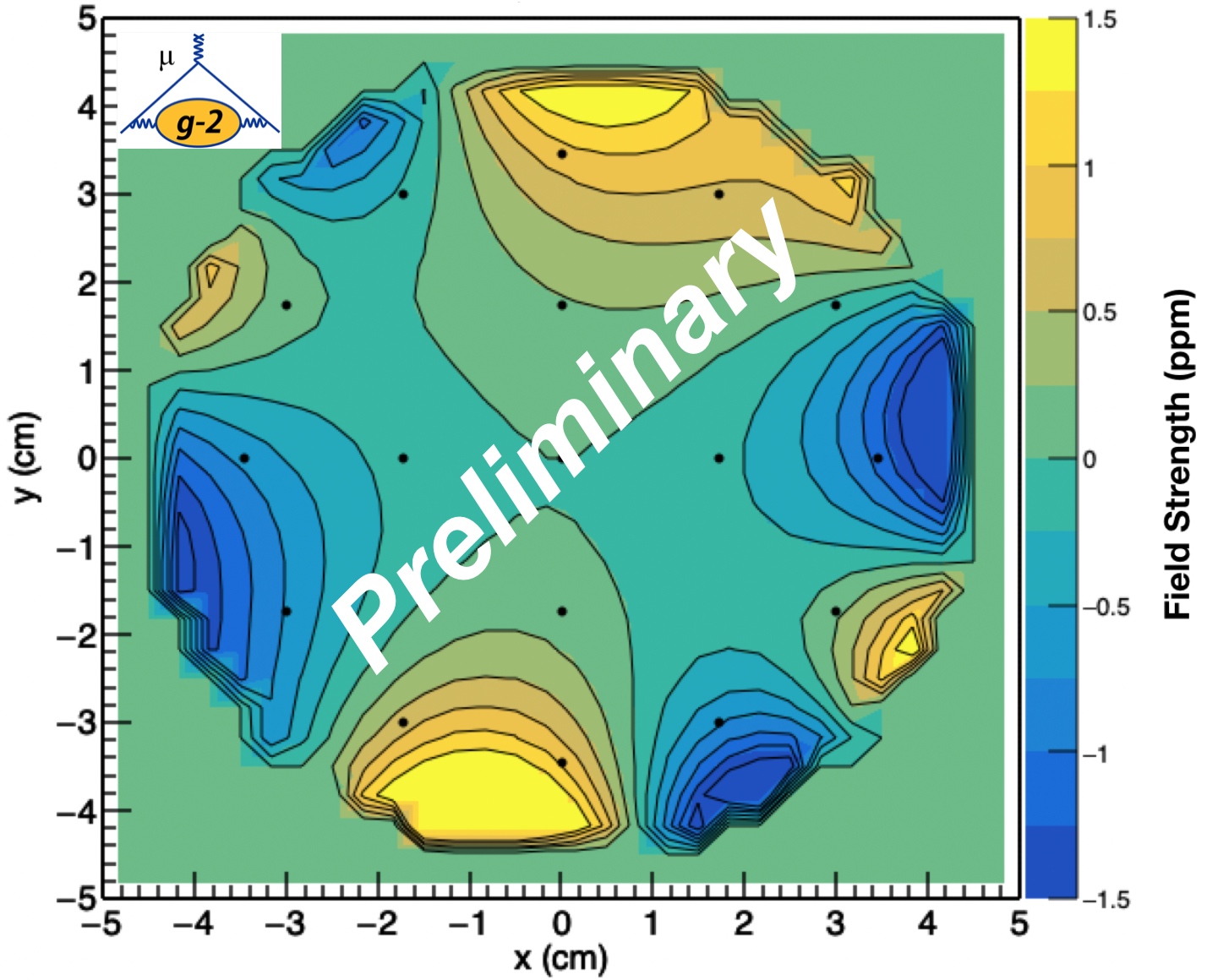} 
    \vspace{-0.2cm}
    \caption{A cross-section of the field in the storage region at one azimuthal location.}
    \label{fig:field}
\end{figure}

The magnetic field is measured by the trolley, which contains 17 NMR probes and completes a full revolution around the ring approximately every three days to map the field profile in the storage region, as shown in \cref{fig:field} for one azimuthal location. The field is constantly measured by the fixed probes located outside of the storage region, in order to track the field between trolley runs. This field map is then convoluted with the beam profile (c.f. \cref{fig:beam}) measured by the trackers to find the average field experienced by the muons before decay. 

\section{Internal alignment of the trackers}
\begin{figure}[!ht]
    \centering
    \includegraphics[width=.54\linewidth]{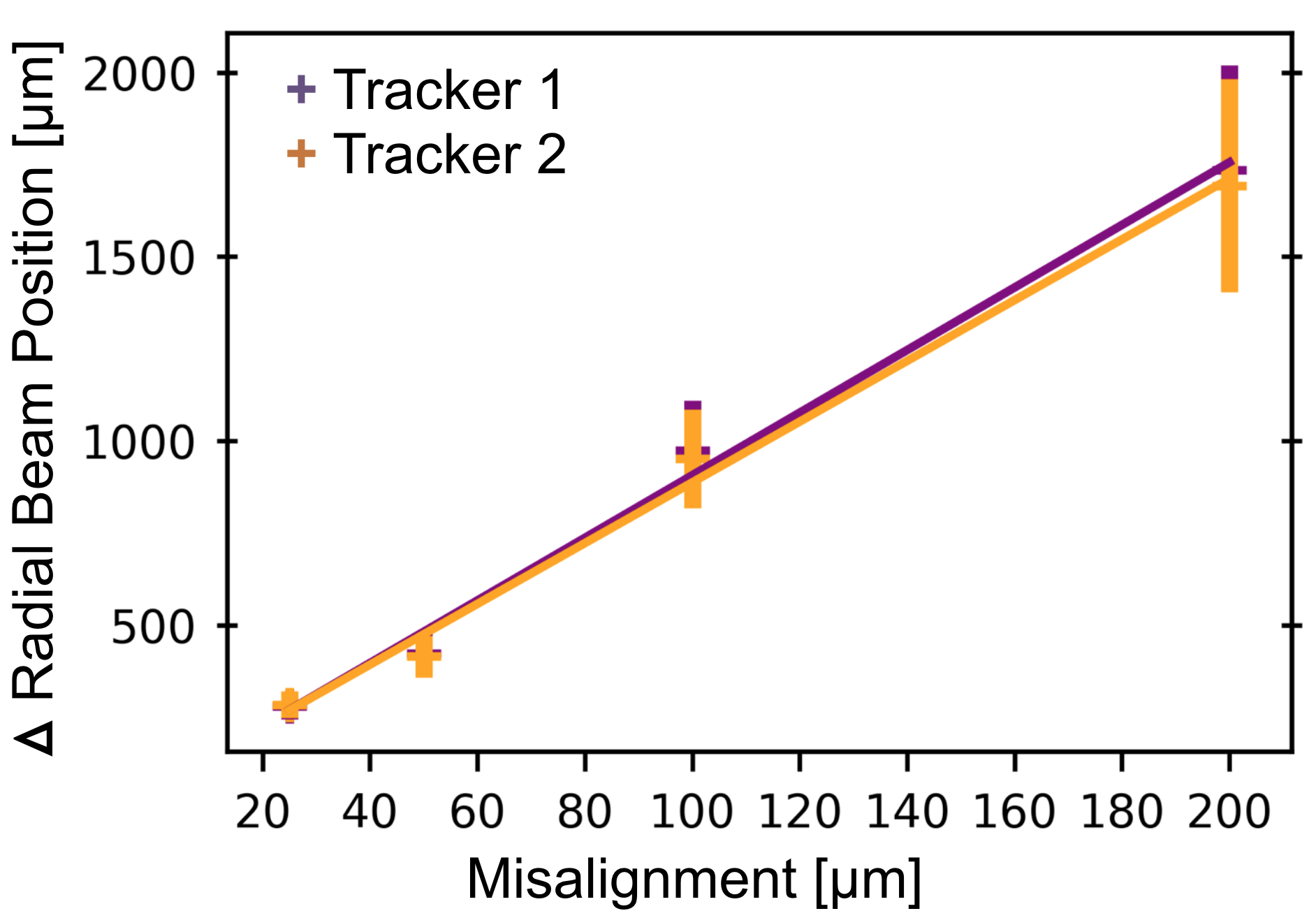} 
    \vspace{-0.25cm}
    \caption{The difference in the radial extrapolated position resulting from a radial misalignment at scales of $20~\micro\metre$, $50~\micro\metre$, $100~\micro\metre$, and $200~\micro\metre$.}
    \label{fig:mis}
\end{figure}

It is essential that the trackers are correctly aligned in order to accurately reconstruct the beam profile, which is necessary to achieve the reduction in the systematic uncertainties required. \cref{fig:mis} shows how an applied randomised uniform radial misalignment at scales of $20~\micro\metre$, $50~\micro\metre$, $100~\micro\metre$, and $200~\micro\metre$ affects the radial extrapolated position of the beam in both stations.

Track-based alignment was implemented with data from Run-1 using the \verb!Millepede-II! framework \cite{Blobel, Blobel2}, minimising $\chi^2$ as a function of track, $\boldsymbol{b}$, and geometry, $\boldsymbol{a}$, parameters 
\begin{equation}
    \chi^2(\boldsymbol{a}, \boldsymbol{b}) = \sum_{j}^{tracks}\sum_{i}^{hits}  \frac{\big(r_{i,j}(\boldsymbol{a},\boldsymbol{b}_j)\big)^2}{(\sigma^{\mathrm{det}})^2},
    \label{eq:mp2}
\end{equation}
where $r$ is the residual, defined as the difference between the fitted track and the measurement in the straw (see \cref{fig:straw}), and $\sigma^{\mathrm{det}}$ is the detector resolution. 
\clearpage
\vspace{-0.05cm}
\verb!Millepede-II! performs a linear squares fit on \cref{eq:mp2}, and returns corrections to the geometry parameters (e.g.~radial and vertical translations). These corrections are added to the assumed position of the modules, and the tracks are then fitted through the aligned detector. A comparison of the tracks before and after the alignment can then be performed. Tracks through the aligned detector now have reduced residuals, due to the correct module placement being used in the track reconstruction. 

\begin{figure}[!ht]
    \centering
    \includegraphics[width=.85\linewidth]{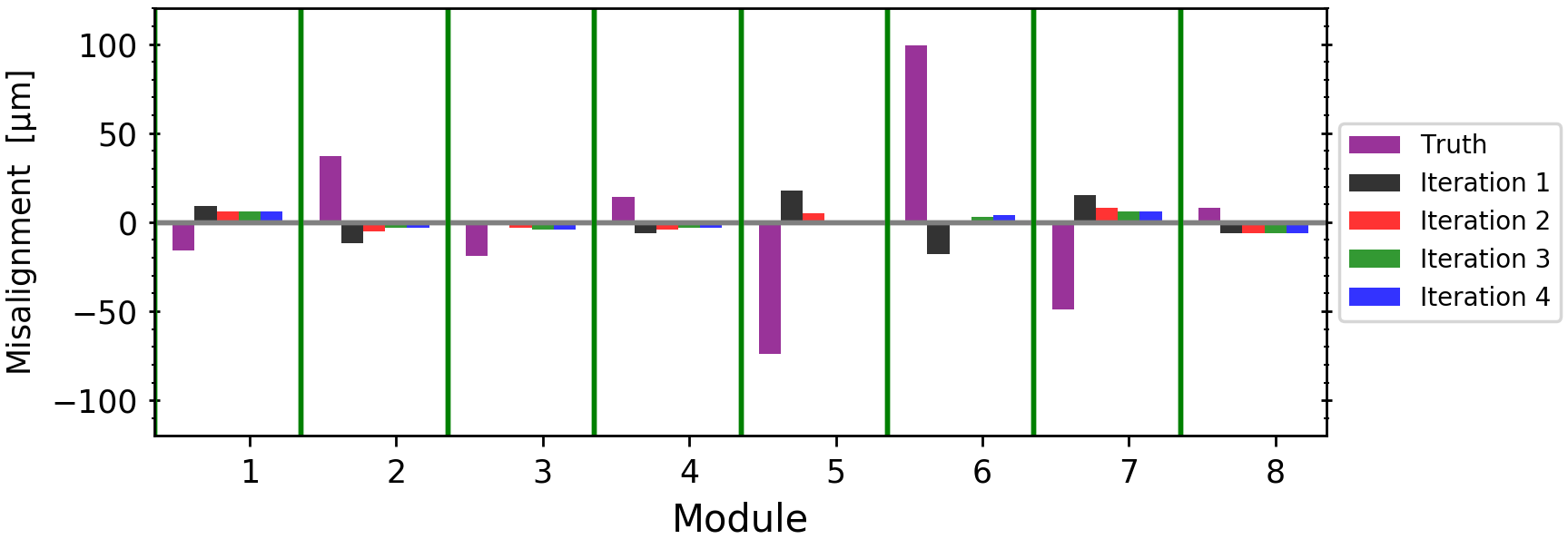}
    \vspace{-0.25cm}
    \caption{Radial alignment results per module in a tracker station in simulation. The truth (input) misalignment is shown in purple. The results of the reconstructed misalignment, for the four iterations, are shown as the difference between the truth and a reconstructed misalignment for an iteration. The reconstructed misalignment from iterations 3 and 4 overlap, indicating convergence, and are shown in green and blue, respectively.}
     \label{fig:sim}
\end{figure}

Alignment convergence in simulation was reached within $2~\micro\metre$ radially and $10~\micro\metre$ vertically, on average per module. Simulation results after four iterations with $\mathcal{O}$($10^5$) tracks are shown in \cref{fig:sim}.
\vspace{-0.05cm}

\begin{figure}[!ht]
    \centering
    \subfigure[]{\raisebox{10mm}{\includegraphics[width=.49\linewidth]{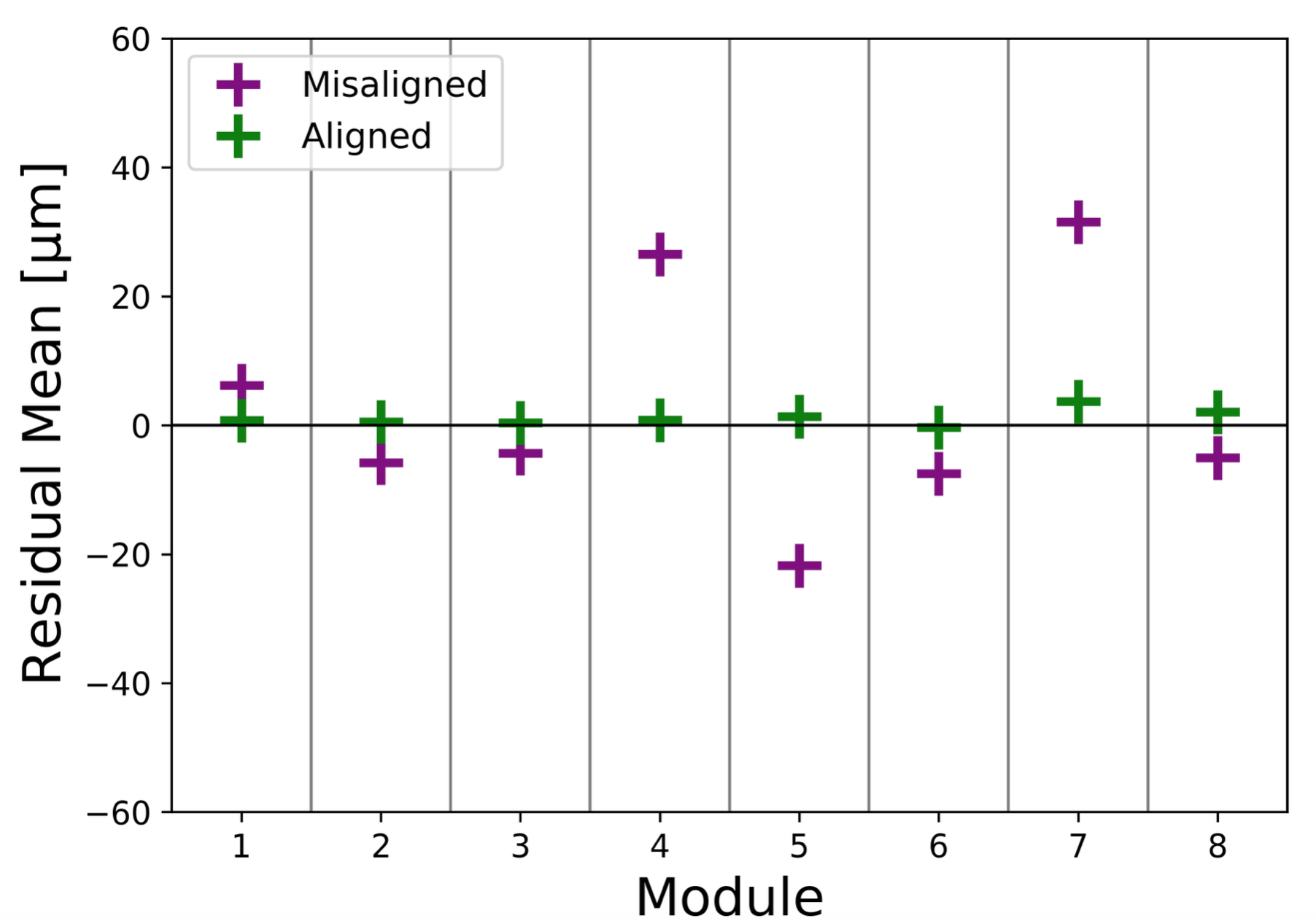} \label{fig:uv}}}
    \subfigure[]{\includegraphics[width=.48\linewidth]{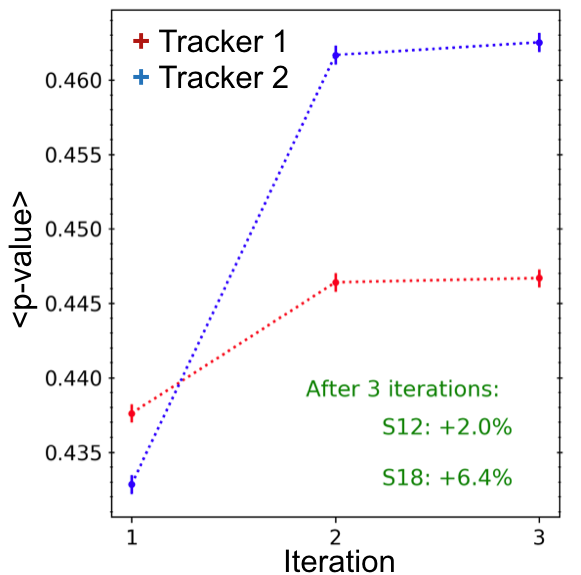} \label{fig:pval}}
    \caption{Alignment results with data: a) The mean residual per module before (purple) and after (green) the alignment in one tracker station. b) The mean p-value of the fitted tracks as a function of the alignment iteration number in both stations.}
\end{figure}

Once the alignment was validated using simulation, the alignment was run over the reconstructed tracks from Run-1. The improvement in the residuals is shown in \cref{fig:uv}, which summarises the distribution of residuals in a tracker station before and after the alignment. The number of reconstructed tracks has increased by 6\% due to the position calibration from the alignment. Moreover, extrapolated tracks have a radial shift towards the centre of the ring of $0.50$~mm and a vertical shift of $0.14$~mm due to the alignment. 

\begin{figure}[!ht]
    \centering
    \includegraphics[width=.85\linewidth]{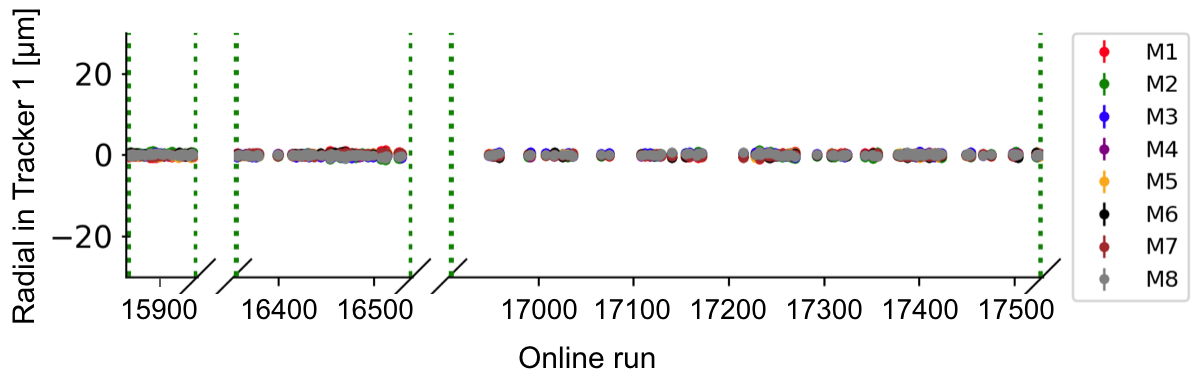} 
    \vspace{-0.3cm}
    \caption{Relative radial alignment results in the eight modules of one tracker station during Run-1, plotted against the online run number. The alignment results are displayed relative to the first online run.}
    \label{fig:mon}
\end{figure}

After the alignment, the uncertainty contribution from the tracker misalignment to the pitch correction is now negligible. Applying the correct alignment to the tracker modules increased the mean p-value of the fitted tracks, as shown in \cref{fig:pval}. The alignment stability was also monitored throughout the entirety of Run-1, as shown in \cref{fig:mon}.
\vspace{-0.1cm}
\vspace{-0.3cm}
\vspace{-0.2cm}
\section{Electric dipole moment}
Analogous to $\boldsymbol{\mu}$ in \cref{eq:mdm}, a potential muon EDM, $\boldsymbol{d_{\mu}}$, is defined by 
\begin{equation}
\boldsymbol{d_{\mu}}=\eta \left(\frac{e}{2m_{\mu}}\right)\boldsymbol{s},
\end{equation}
where $\eta$ is a dimensionless constant \cite{FNAL_TDR}, analogous to $g$ in \cref{eq:mdm}. 

\begin{figure}[ht!]
    \centering
    \includegraphics[height=5.1 cm]{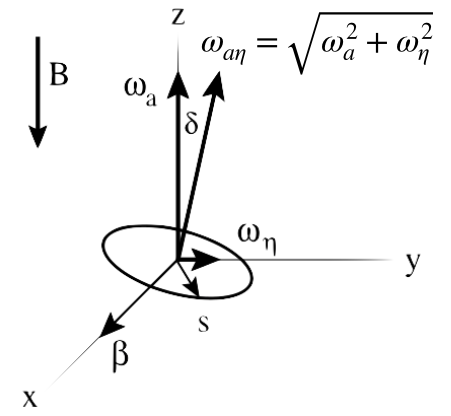}
    \vspace{-0.1cm}
    \caption{Tipping of the precession plane away from the momentum axis ($x$), due to the EDM.}
    \label{fig:edmgeom}
\end{figure}

The trackers can look for a potential muon EDM, which adds an additional term, $\boldsymbol{\omega_{\eta}}$, to the observed precession frequency in \cref{eq:mub} \cite{Saskia}
\begin{equation}
\boldsymbol{\omega_{a\eta}}=\boldsymbol{\omega_{a}}+\boldsymbol{\omega_{\eta}}=a_{\mu}\frac{e}{m_{\mu}}\boldsymbol{B} + \eta\frac{e}{2m_{\mu}}\left(\frac{E}{c}+\beta\times\boldsymbol{B}\right)=\omega_{a}\sqrt{1+\left(\frac{\eta\boldsymbol{\beta}}{2a_{\mu}}\right)}.
\label{eq:edm}
\end{equation}
An EDM would increase the observed spin precession frequency and tilt the precession plane by an angle $\delta$, given by
\begin{equation}
\delta = \tan^{-1}\left(\frac{\omega_{\eta}}{\omega_a}\right)=\tan^{-1}\left(\frac{\eta\beta}{2a_{\mu}}\right),
\end{equation}
as illustrated in \cref{fig:edmgeom}. The goal is to measure $\delta$ to within $0.4~\micro\radian$ to place a new limit on the muon EDM with a sensitivity of up to $10^{-21}~e\cdot$cm \cite{Becky}, with at least a factor of 10 improvement on the current best limit from the BNL experiment \cite{BNL_EDM}.

\begin{figure}[!ht]
    \centering
    \subfigure[]{{\includegraphics[width=.47\linewidth]{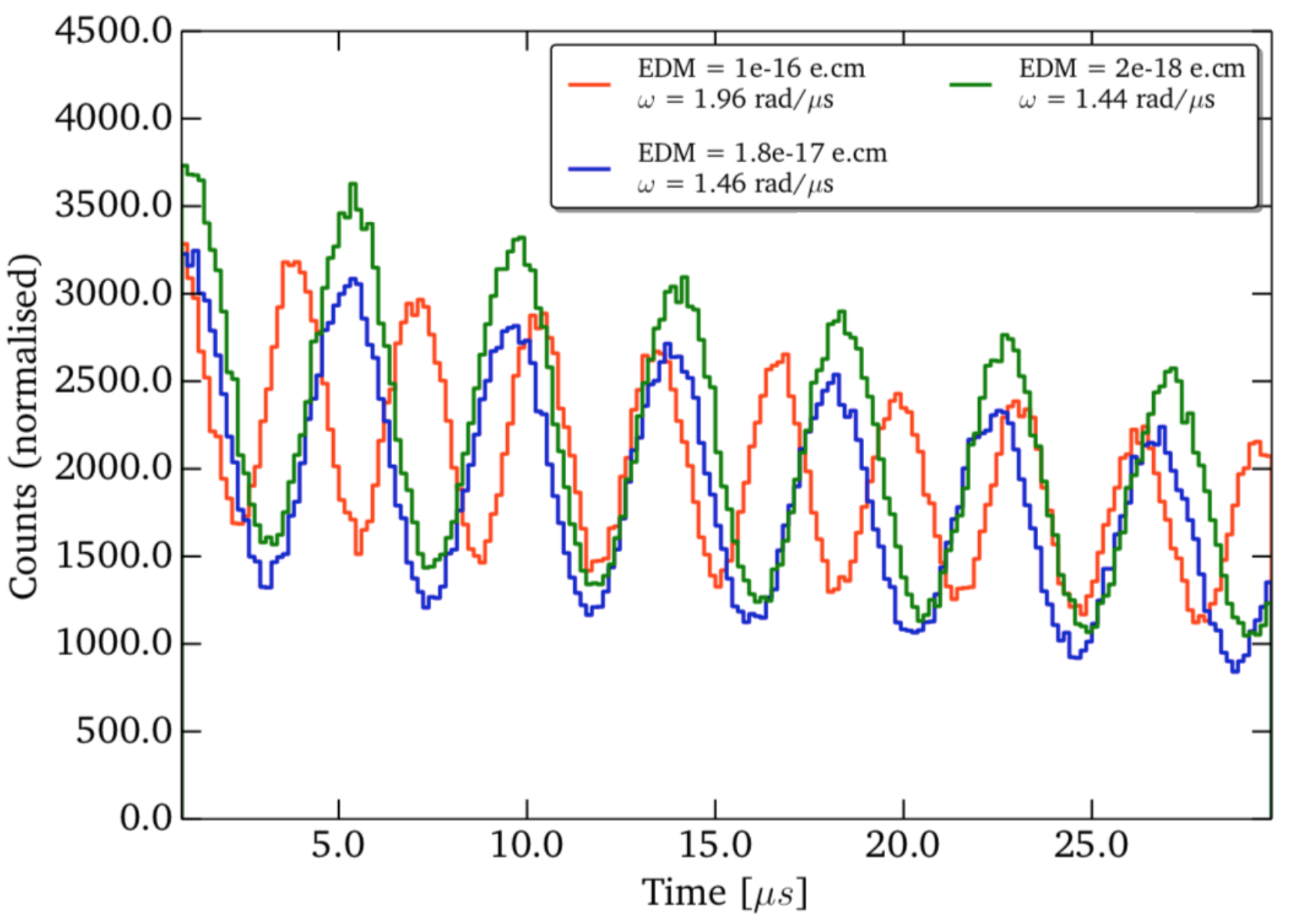} \label{fig:edm1}}}
    \subfigure[]{\includegraphics[width=.47\linewidth]{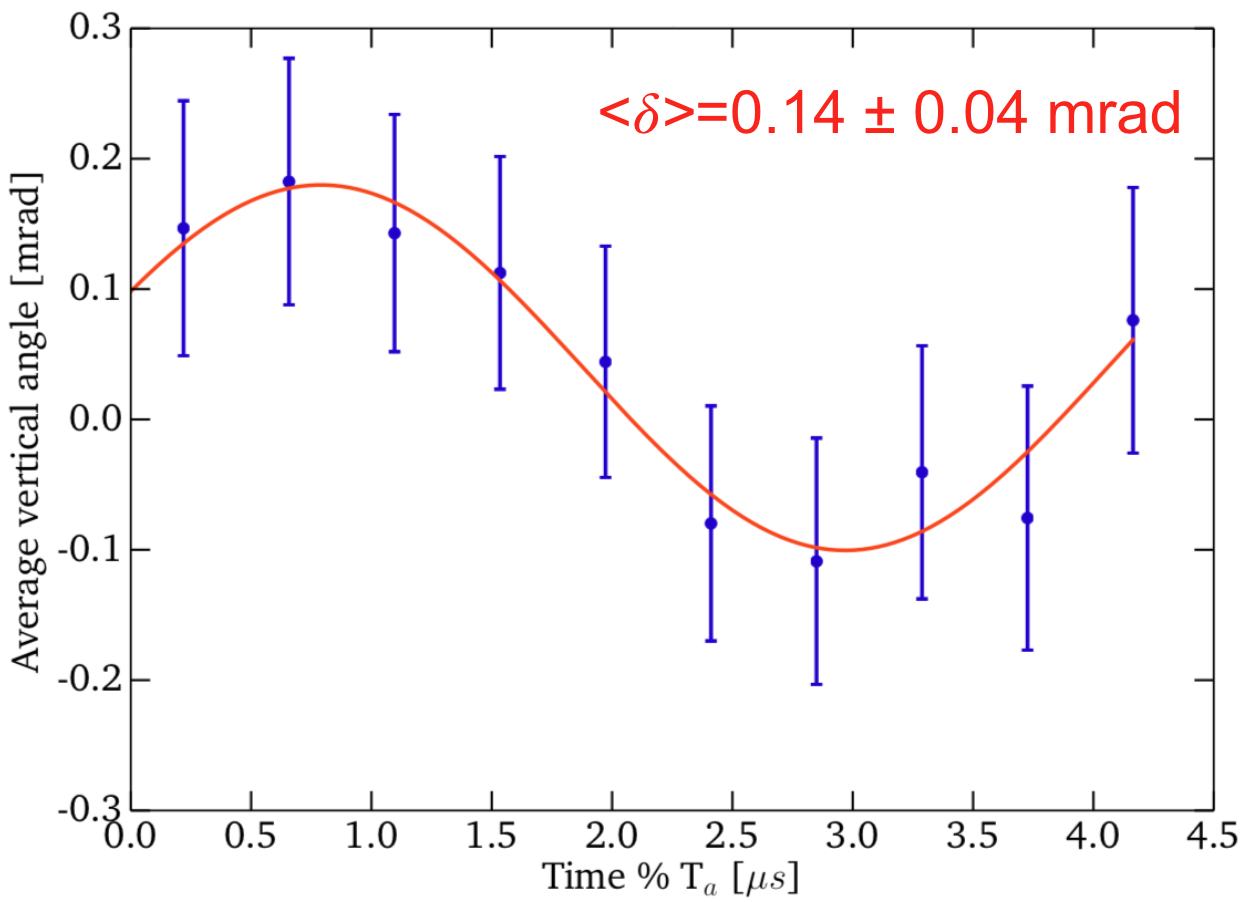} \label{fig:edm2}}
    \vspace{-0.2cm}
    \caption{EDM Simulation results \cite{Saskia}: a) The observed $\omega_{a}$ in the calorimeters (c.f. \cref{fig:wiggle}) for three large input EDM signals. b) The average vertical angle, $\delta$, of the decay positrons seen in the trackers for an input $\boldsymbol{d_{\mu}}=5.4\times10^{-18}~e\cdot$cm.}
\end{figure}

Simulation results \cite{Saskia} that demonstrate how a large input EDM signal changes the observed $\omega_{a\eta}$ in the calorimeters are shown in \cref{fig:edm1}. The oscillation in the average vertical decay angle of the positron is shown in \cref{fig:edm2}.
\vspace{-0.1cm}

The SM limit on $\boldsymbol{d_{\mu}}$ is $10^{-34}~e\cdot$cm \cite{BNL_EDM}, well below the current experimental reach. Hence, any observation of $\boldsymbol{d_{\mu}}$ is evidence of NP. The Hamiltonian for the muon in applied magnetic, $\boldsymbol{B}$, and electric, $\boldsymbol{E}$, fields is given by
\vspace{-0.1cm}
\begin{equation}
\mathcal{H} = -\boldsymbol{\mu}\cdot\boldsymbol{B}-\boldsymbol{d_{\mu}}\cdot\boldsymbol{E}.
\label{eq:H}
\end{equation}
Unlike the magnetic dipole moment term in \cref{eq:H}, the EDM term, $\boldsymbol{d_{\mu}}\cdot\boldsymbol{E}$, is a CP-odd quantity \cite{Becky}. Therefore, it would provide a new source of CP violation in the charged lepton sector.

\section{Conclusions}
The Fermilab $g-2$ experiment is expected to measure the muon magnetic anomaly of the muon to 140~ppb, and will set a new limit on the muon EDM, with at least a factor of 10 improvement on the previous limit. The two data taking periods, Run-1 and Run-2, have been successfully completed, with two more periods anticipated. Essential in adhering to the experimental systematic uncertainty budget are the straw tracking detectors, which perform track extrapolation backwards to the muon decay point and forwards to the calorimeters. The trackers reduce the systematic uncertainty on $\boldsymbol{\omega_a}$ via measurements of the amplitude and frequency of the radial betatron motion, investigations of pileup events, and assessments of the calorimeter gains. Moreover, systematic effects, such as the vertical pitch, require a correction that is accessible via a measurement of the vertical width of the beam. The beam profile from the trackers is also convoluted with the magnetic field map to find the field experienced by the muons at the point of decay. 

In order to accurately reconstruct the beam profile, it is essential that the trackers are correctly aligned. The alignment algorithms were validated using simulation, which converged after three iterations with $\mathcal{O}$($10^5$) tracks to within $2~\micro\metre$ radially and $10~\micro\metre$ vertically. Track-based internal alignment was implemented with data from Run-1. The number of reconstructed tracks has increased by 6\% due to the position calibration from the alignment. After the alignment procedure, the uncertainty contribution from the tracker misalignment to the pitch correction is now negligible.

A potential EDM of the muon would increase the observed $\boldsymbol{\omega_a}$ signal and tilt the precession plane of the muon. The trackers will realise an EDM measurement through the direct detection of an oscillation in the average vertical angle of the decay positron. An observation of a muon EDM would be evidence of NP, and would provide a new source of CP violation in the charged lepton sector.

\vspace*{-5pt}
\section*{Acknowledgements}
The Fermilab Muon $g-2$ experiment is supported in part by the US Department of Energy, the US National Science Foundation, the Istituto Nazionale di Fisica Nucleare in Italy, the UK Science and Technology Facilities Council, the UK Royal Society, the European Union’s Horizon 2020 research and
innovation programme under the Marie Sk\l{}odowska-Curie grant agreements
No.~690835 (MUSE), No.~734303 (NEWS), MOST and NSFC in China, MSIP and NRF in Korea.\\
The author is supported in part by the Visiting Scholars Award Program of the Universities Research Association.

\end{document}